\documentclass[aps,pra,showpacs,groupedaddress,twocolumn]{revtex4}
\usepackage{eurosym}
\usepackage{amsfonts}
\usepackage{graphicx}
\usepackage{latexsym}
\usepackage{makeidx}
\usepackage{amsmath}
\usepackage{amssymb}

\begin{document}

\title{Cavity Dynamical Casimir Effect in the presence of three-level atom}
\author{A. V. Dodonov and V. V. Dodonov}
\affiliation{Instituto de F\'{\i}sica, Universidade de Bras\'{\i}lia, PO Box 04455,
70910-900, Bras\'{\i}lia, Distrito Federal, Brazil}

\begin{abstract}
We consider the scenario in which a damped three-level atom in the ladder or
V configurations is coupled to a single cavity mode whose vacuum state is
amplified by dint of the dynamical Casimir effect. We obtain approximate
analytical expressions and exact numerical results for the time-dependent
probabilities, demonstrating that the presence of the third level modifies
the photon statistics and its population can serve as a witness of photon
generation from vacuum.
\end{abstract}

\maketitle

\section{Introduction}

The physics of three-level quantum systems (\textquotedblleft
atoms\textquotedblright ) interacting with quantized modes of
electromagnetic field is very rich, and many special cases were studied in
numerous papers (see, e.g., \cite%
{KnightMil80,YooEb,Chum88,Buz90,GerEb90,Shum91,Alex95,Wu97,Xu99,Klim00,Luk02,Mess03}
and references therein). In the majority of studies, the coefficients of the
Hamiltonians describing such systems were assumed time-independent.
Time-dependent couplings were considered, e.g., in \cite{Janow03}, but under
the restriction of adiabatic variations. Here we consider the light-matter
dynamics when a three-level atom interacts with a single cavity mode, whose
frequency is rapidly modulated according to the harmonical law $\omega
_{t}=\omega _{0}[1+\varepsilon \sin (\eta t)]$ with a small modulation
depth, $|\varepsilon |\ll 1$. We shall use dimensionless variables, setting $%
\hbar =\omega _{0}=1$. Such a situation can arise, in particular, if the
selected mode describes the evolution of electromagnetic field in a cavity
with vibrating walls, and one of the most impressive manifestations can be
the so called Dynamical Casimir Effect (DCE), i.e., the photon generation
from the initial vacuum state induced by the motion of boundaries \cite%
{revDCE,revDal}. The simplest model describing this effect is based on the
Hamiltonian
\begin{equation}
H_{c}=\omega _{t}n-i\chi _{t}\left( a^{2}-a^{\dagger 2}\right) ,  \label{Hc}
\end{equation}
where $a$ and $a^{\dagger }$ are the cavity annihilation and creation
operators, and $n\equiv a^{\dagger }a$ is the photon number operator. The
specific feature of the DCE is that two functions $\omega _{t}$ and $\chi_{t}
$ are related as follows \cite{Law94}:
\begin{equation}
\chi_{t}=(4\omega _{t})^{-1}d\omega _{t}/dt,  \label{chit}
\end{equation}%
If the modulation frequency $\eta $ is close to the parametric resonance
frequency, $\eta =2(1+x)$ with $|x|\ll 1$, then one can expect an
exponential growth of the mean number of photons inside the \emph{empty
ideal\/} cavity \cite{DK96}. In particular, the mean number of photons
created from the initial vacuum state for $x=0$ equals
\begin{equation}
\langle n_0(t)\rangle = \sinh^2\left(\varepsilon t/2\right).  \label{n0}
\end{equation}

However, the situation can be very different if the field mode interacts
with a detector while the cavity walls oscillate. For example, it was shown
in \cite{PLA} that no more than \emph{two photons\/} can be created in the
cavity if the field-atom coupling is much stronger than the modulation depth
$\varepsilon$. In view of the recent progress in experiments on simulating
DCE \cite{DCE-Nature,revRMP,Agnesi11}, the detailed study of different
detection schemes becomes a timely and important task.

\begin{figure}[tbh]
\begin{center}
\vspace{-7mm}\includegraphics[width=0.49\textwidth]{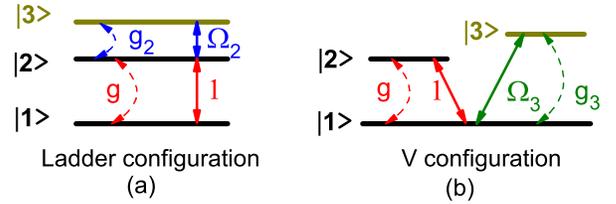} {}
\end{center}
\par
\vspace{-1cm}
\caption{(Color online) Diagram of the atomic configurations.}
\label{f1}
\end{figure}

Recently, various regimes of the \emph{two-level\/} detectors were analyzed
in \cite{Roberto,Resdyn,Resact}. On the other hand, \emph{three-level\/}
detectors can be more realistic \cite{Pero11}, besides, they have several
advantages \cite{Pinot08}. Therefore we consider the evolution of the
single-mode cavity field interacting with a three-level atom, whose free
Hamiltonian is
\begin{equation}
H_{a}=\tilde{E}_{1}\sigma _{11}+\tilde{E}_{2}\sigma _{22}+\tilde{E}%
_{3}\sigma _{33},
\end{equation}
(where $\tilde{E}_{i}$ is the $i$th energy level, $|\mathbf{i}\rangle $ is
the atomic energy eigenstate and $\sigma _{ij}\equiv |\mathbf{i}\rangle
\langle \mathbf{j}|$). We are interested in the cases where the atom-field
interaction modifies severely the atomless DCE. Therefore we assume that the
$|\mathbf{1\rangle }\leftrightarrow |\mathbf{2\rangle }$ transition is
resonant with the unperturbed cavity frequency, $\tilde{E}_{2}=\tilde{E}%
_{1}+1$. Fig. \ref{f1} depicts two atomic level structures we consider here:
the ladder (or $\Xi $) configuration and the V configuration, where $%
\Omega _{2}\equiv \tilde{E}_{3}-\tilde{E}_{2}$ and $\Omega _{3}\equiv $ $%
\tilde{E}_{3}-\tilde{E}_{1}$ are the transition frequencies.
We also define the detunings between the cavity unperturbed frequency and
the other atomic transition frequency as $\Delta _{2}\equiv 1-\Omega _{2}$
and $\Delta _{3}\equiv 1-\Omega _{3}$. The set of constants $g$, $g_{2}$ and
$g_{3}$ quantifies the atom-field dipolar coupling strengths between the
energy levels as shown in Fig. \ref{f1}. We assume these constants to be
real and $\mathcal{O}(g_{1})\sim \mathcal{O}(g_{2})\sim \mathcal{O}(g_{3})$
(although this is not the most general case, such a choice describes the
main phenomena in the most simple way) and much smaller than unity. The
corresponding light-matter interaction Hamiltonians are chosen in the
standard Jaynes--Cummings form (i.e., we neglect the counter-rotating
terms):
\begin{equation}
H_{I}^{(\Xi)}=a(g\sigma _{21}+g_{2}\sigma _{32})+H.c.,  \label{HIlad}
\end{equation}
\begin{equation}
H_{I}^{(V)}=a(g\sigma _{21}+g_{3}\sigma _{31})+H.c.,  \label{HIV}
\end{equation}
where $H.c.$ stands for the Hermitian conjugate.

The questions we try to answer are: 1) how the presence of the third level
can influence the number of created photons and the photon number
distribution and 2) how big can the occupation probabilities of different
levels be (this is important from the point of view of the detection of
DCE). For this purpose we solved numerically the Schr\"{o}dinger equation
\begin{equation}
d|\Psi \rangle /dt=-i\left(H_{c}+H_{a}+H_{I}\right)|\Psi \rangle
\label{SEtot}
\end{equation}
for the wavefunction of the total system, expanding this function over the
atomic  and Fock bases. Exact equations for the coefficients of this
expansion were solved numerically using the Runge-Kutta-Verner fifth-order
and sixth-order method, truncating the photon number space at the value $%
N=196$. In this paper we focus on the amplification of the vacuum
fluctuations, so we consider only the zero-excitation initial state, $|\Psi
(0)\rangle =|\mathbf{1},0\rangle $.

To be closer to realistic experimental conditions, we took into account in
some cases (where the mean number of created photons was limited) a
possibility of damping in the atomic degrees of freedom (but neglecting the
dissipation in the field mode, assuming that the cavity quality factor is
high enough), using the Lindblad-type Markovian master equation for the
total statistical operator $\rho $ of the atom-field system (in the ladder configuration)
\begin{equation}
\dot{\rho}=-i\left[ H,\rho \right] +\lambda \mathfrak{D}\left[ \sigma _{12}%
\right] \rho +\lambda _{2}\mathfrak{D}[\sigma _{23}]\rho ,  \label{meq}
\end{equation}%
where $\lambda$ and $\lambda _{2}$ are the damping rates for the transitions
$|\mathbf{2}\rangle \rightarrow |\mathbf{1}\rangle $ and $|\mathbf{3}\rangle
\rightarrow |\mathbf{2}\rangle $, respectively, and the Lindblad kernel is
\begin{equation*}
\mathfrak{D}[O]\rho \equiv (2O\rho O^{\dagger }-O^{\dagger }O\rho -\rho
O^{\dagger }O)/2~.
\end{equation*}%
The maximal number of photons taken into account in such cases was $N=7$ due
to the necessity to calculate off-diagonal matrix elements not only in the
atomic basis but also in the Fock one.  We checked that the normalization
conditions were fulfilled with an accuracy better than $10^{-10}$ in all the
cases.

The results of numerical calculations are exposed in the next two sections
together with some approximate analytical solutions clarifying them. We
consider two typical situations: the strong atom--field coupling regime with
$|g|\gg |\varepsilon |$ and the weak coupling regime with $|g|\ll
|\varepsilon |$. Both these regimes could be implemented in the circuit QED
realizations, where the values of $g$ can be adjusted from very low values
up to $|g|\sim 10^{-1}$ during fabrication or \emph{in situ} \cite{Koch}.
The last section contains some discussion and conclusions.

\section{Ladder configuration}

\subsection{Main resonance for a strong field-atom coupling}

\begin{figure}[htb]
\begin{center}
\includegraphics[width=0.49\textwidth]{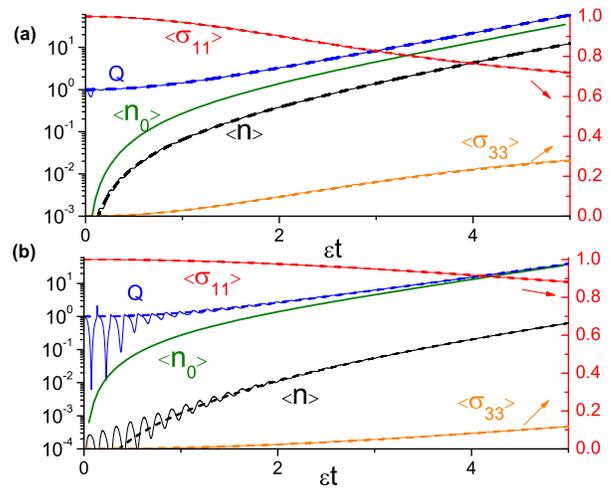} {}
\end{center}
\par
\vspace{-1cm}
\caption{(Color online) \textbf{a)} The behavior of different average values
(described in the text) versus the dimensionless time $\protect\varepsilon t$
in the strong modulation regime for $x=0$ and the following values of other
parameters: \textbf{a) } $\protect\lambda =\Delta _{2}=0$, $g=3\times 10^{-2}
$, $g_{2}=4\times 10^{-2}$, $\protect\varepsilon =10^{-3}$; \textbf{b) } the
same as (a) but for $g_{2}=10^{-2}$.}
\label{fig2new}
\end{figure}

In Fig. \ref{fig2new} we demonstrate the behavior of typical quantities
characterizing the field and atom dynamics: the average photon number $%
\left\langle n\right\rangle $, the atomic level populations $\sigma _{ii}$ ($%
i=1,2,3$) and the Mandel factor $Q=[\left\langle (\Delta n)^{2}\right\rangle
-\left\langle n\right\rangle ]/\left\langle n\right\rangle $, for the
resonance shift $x=0$ and resonant third level in the absence of any damping
and under the condition of strong field-atom coupling: $|\varepsilon| \ll
|g|, |g_2|$. This case is especially interesting, because it gives the
maximal photon generation rate for the empty cavity \cite{DK96}. On the
other hand, according to \cite{PLA}, there is no photon creation for $x=0$
if the field mode interacts with a \emph{two-level\/} atom under the
condition $|\varepsilon |\ll |g|$.

Solid lines correspond to exact numerical results and the dashed ones -- to
the approximation based on Eqs. (\ref{er1})-(\ref{er2}) deduced  below. We
see that for $\varepsilon t>1$ the mean number of photons grows
exponentially with the same increment $d\ln\langle n\rangle/dt$ as in the
empty-cavity case described by Eq. (\ref{n0}), although $\langle n\rangle$
can be much less than $\langle n_0\rangle$ if $g_2 <g$.

To understand qualitatively how the coupling with the third level changes
the system dynamics we used the following chain of approximations.  In the
weak modulation case $|\varepsilon |\ll 1$ considered here, one can write $%
\chi _{t}\simeq 2q\cos (\eta t)$ with $q\equiv \varepsilon (1+x)/4$. Going
to the interaction picture via the transformation $|\Psi (t)\rangle
=V(t)|\psi (t)\rangle $, with $V\left( t\right) =\exp [-it\left( \eta
/2\right) (n+\sigma _{33}-\sigma _{11})]$, and performing the Rotating Wave
Approximation (RWA), one obtains the approximate \emph{time-independent\/}
Hamiltonian governing the time evolution of $|\psi (t)\rangle $%
\begin{eqnarray}
H_{1} &\simeq &(ga\sigma _{21}+g_{2}a\sigma _{32}-iqa^{2}+H.c.)  \notag \\
&&+x(\sigma _{11}-n-\sigma _{33})-\Delta _{2}\sigma _{33}.  \label{H1}
\end{eqnarray}%
It is convenient to expand the function $|\psi (t)\rangle $ over the atomic
and Fock basis as follows:
\begin{eqnarray}
|\psi (t)\rangle &=&\sum_{n=0}^{\infty }e^{inxt}\left( e^{-itx}p_{\mathbf{1}%
n}|\mathbf{1},n\rangle +p_{\mathbf{2}n}|\mathbf{2},n\rangle \right.  \notag
\\
&&\left. +e^{it\left( x+\Delta _{2}\right) }p_{\mathbf{3}n}|\mathbf{3}%
,n\rangle \right),  \label{psi}
\end{eqnarray}%
so that $|p_{\mathbf{i}n}|^{2}$ is the
probability of the atom to be in the $\mathbf{i}-$th state with the field
having $n$ photons.
The time-dependent phase factors are introduced here to simplify
the equations for the coefficients $p_{\mathbf{i}n}$. Putting (\ref{psi})
into the equation $d|\psi \rangle /dt=-iH_{1}|\psi \rangle $ we obtain the
following differential equations for the probability amplitudes:%
\begin{equation}
\dot{p}_{\mathbf{1}n}=-i\sqrt{n}gp_{\mathbf{2}\left( n-1\right) }+\mathcal{W}%
_{\mathbf{1}}(n)  \label{line1}
\end{equation}%
\begin{equation}
\dot{p}_{\mathbf{2}n}=-ig\sqrt{n+1}p_{\mathbf{1}\left( n+1\right) }-ig_{2}%
\sqrt{n}e^{i\Delta _{2}t}p_{\mathbf{3}\left( n-1\right) }+\mathcal{W}_{%
\mathbf{2}}(n)  \label{line2}
\end{equation}%
\begin{equation}
\dot{p}_{\mathbf{3}n}=-ig_{2}\sqrt{n+1}e^{-i\Delta _{2}t}p_{\mathbf{2}\left(
n+1\right) }+\mathcal{W}_{\mathbf{3}}(n),  \label{line3}
\end{equation}%
where for $\mathbf{i}=\mathbf{1},\mathbf{2},\mathbf{3}$%
\begin{eqnarray}
\mathcal{W}_{\mathbf{i}}(n) &\equiv &q\Big[\sqrt{n(n-1)}e^{-2ixt}p_{\mathbf{i%
}(n-2)}  \notag \\
&&-\sqrt{(n+1)(n+2)}e^{2ixt}p_{\mathbf{i}(n+2)}\Big].  \label{Wi}
\end{eqnarray}%
Analyzing Eqs. (\ref{line1})-{\ref{Wi}) one can see that for the initial
state $|\mathbf{1},0\rangle $ the only nonzero coefficients at $t>0$ can be $%
p_{\mathbf{1}\left( 2k\right) }$, $p_{\mathbf{2}\left( 2k+1\right) }$, and $%
p_{\mathbf{3}\left( 2k\right) }$, with $k=0,1,2,\ldots$. }

For $|\varepsilon |\ll |g|$ we follow the scheme used in \cite{PLA,Roberto}.
First we solve Eqs. (\ref{line1})-(\ref{line3}) for $q=0$ (i.e., in the
stationary cavity). In the strict resonant case, $\Delta _{2}=0$, we have  $%
p_{\mathbf{1}0}(t)=\mathrm{const}$, whereas for $n= 2,4,6,\ldots$  the
solutions can be written as follows:
\begin{equation}
p_{\mathbf{1}n}=\frac{g\sqrt{n}}{G_{n}}\left(
A_{n}e^{-iG_{n}t}-B_{n}e^{iG_{n}t}-C_{n}\right) ,  \label{li1}
\end{equation}%
\begin{equation}
p_{\mathbf{3}\left( n-2\right) }=\frac{g_{2}\sqrt{n-1}}{G_{n}}\left[
A_{n}e^{-iG_{n}t}-B_{n}e^{iG_{n}t}+\frac{ng^{2}C_{n}}{(n-1)g_{2}^{2}}\right]
\label{li2}
\end{equation}%
\begin{equation}
p_{\mathbf{2}\left( n-1\right) }=A_{n}e^{-iG_{n}t}+B_{n}e^{iG_{n}t},
\label{li3}
\end{equation}%
where $G_{n}\equiv \sqrt{ng^{2}+\left( n-1\right) g_{2}^{2}}$ and constants $%
A_{n},B_{n},C_{n}$ are determined by the initial conditions. For $q \neq 0$
we put expressions (\ref{li1})-(\ref{li3}) into Eqs. (\ref{line1})-(\ref%
{line3}), obtaining the equivalent equations $\dot{p}_{\mathbf{1}0}=-q\sqrt{2%
}e^{2ixt}p_{\mathbf{1}2}$ and for $n\geq 2$
\begin{equation}
\dot{A}_{n}e^{-iG_{n}t}-\dot{B}_{n}e^{iG_{n}t}-\dot{C}_{n}=\frac{G_{n}}{g%
\sqrt{n}}\mathcal{W}_{\mathbf{1}}(n)  \label{q1}
\end{equation}%
\begin{equation}
\dot{A}_{n}e^{-iG_{n}t}-\dot{B}_{n}e^{iG_{n}t}+\frac{ng^{2}\dot{C}_{n}}{%
\left( n-1\right) g_{2}^{2}}=\frac{G_{n}\mathcal{W}_{\mathbf{3}}(n-2)}{g_{2}%
\sqrt{n-1}}  \label{q2}
\end{equation}%
\begin{equation}
\dot{A}_{n}e^{-iG_{n}t}+\dot{B}_{n}e^{iG_{n}t}=\mathcal{W}_{\mathbf{2}}(n-1).
\label{q3}
\end{equation}%
According to numerical results, all coefficients $p_{\mathbf{2} n }$ are
very small for $x=0$. Therefore we neglect functions $A_{n}(t)$  and $%
B_{n}(t)$ in all terms, except for the left-hand sides of Eqs. (\ref{q1})
and (\ref{q2}), because the derivatives $\dot{A}_{n}$ and $\dot{B}_{n}$  can
be not small due to fast oscillations of these functions with the
frequencies  of the order of $G_{n}$. To eliminate these derivatives we take
the difference of equations (\ref{q1})  and (\ref{q2}), arriving at the set
of equations containing only coefficients  $C_{n}$ --- here we make the
approximation, removing $A_{n}(t)$  and $B_{n}(t)$ from the terms $\mathcal{W%
}_{\mathbf{1}}(n)$ and $\mathcal{W}_{\mathbf{3}}(n)$.  After that we rewrite
the coefficients $C_{n}$ in terms of $p_{\mathbf{1}n}$ according  to Eq. (%
\ref{li1}) with neglected $A_{n}(t)$ and $B_{n}(t)$. Thus we obtain the
following infinite set of differential equations coupling the functions ${p}%
_{\mathbf{1}n}$ only:
\begin{equation}
\dot{p}_{\mathbf{1}0}= -(\varepsilon/4)\sqrt{2} {p}_{\mathbf{1}2},
\label{er1}
\end{equation}
\begin{equation}
\dot{p}_{\mathbf{1}2}= (\varepsilon/4)\left[ \frac{g_{2}^{2}\sqrt{2}}{G_{2}^2%
}p_{\mathbf{1}0} - \frac{2\tilde{G}_{2}^{2}}{\sqrt{3}G_{2}^2} {p}_{\mathbf{1}%
4} \right] ,  \label{er12}
\end{equation}
\begin{eqnarray}
\dot{p}_{\mathbf{1}n} &=& (\varepsilon/4)(n-1)\left[ \sqrt{\frac{n}{n-1}}%
\frac{\tilde{G}_{n-2}^{2}}{G_{n}^2}{p}_{\mathbf{1}(n-2)} \right.  \notag \\
&&\left. - \sqrt{\frac{n+2}{n+1}}\frac{\tilde{G}_{n}^{2}}{G_{n}^2} {p}_{%
\mathbf{1}(n+2)} \right] ,  \label{er2}
\end{eqnarray}%
where $\tilde{G}_{n}\equiv \sqrt{ng^{2}+\left( n+1\right) g_{2}^{2}}$ and $n
= 4,6,\ldots$. The amplitudes ${p}_{\mathbf{3}n}$ can be calculated by means
of the relation
\begin{equation}
p_{\mathbf{3}n} \simeq -(g/g_{2})\sqrt{(n+2)/(n+1)}p_{\mathbf{1}\left(
n+2\right)},  \label{vnb}
\end{equation}%
which follows from Eq. (\ref{line2}) if one puts $p_{\mathbf{2}n}\approx 0$
there.

Although equations (\ref{er1})-(\ref{er2}) cannot be solved analytically due
to the presence of various square roots in the coefficients, they are very
useful, both for numerical calculations and the qualitative analysis. Fig. %
\ref{fig2new} shows that differences between exact solutions of the full
Schr\"odinger equation (\ref{SEtot}) and the approximate ones based on the
set (\ref{er1})-(\ref{er2}) practically disappear in the most interesting
regime $\varepsilon t >1$. But solving (\ref{er1})-(\ref{er2}) numerically
requires much less resources than solving (\ref{SEtot}), therefore using (%
\ref{er1})-(\ref{er2}) we can calculate the amplitudes for much bigger
values of the dimensionless time $\varepsilon t $. In this way we confirmed
numerically that the exponential growth of the mean photon number continues (at least for
$\varepsilon t \lesssim 10$). The population of the second level
shows fast oscillations, not exceeding the values of the order of $%
(\varepsilon/g)^2$, as can be evaluated from Eqs. (\ref{li3}) and (\ref{q1}).

The order of magnitude of amplitudes $p_{\mathbf{1}n}$ with $n \ge 2$ is
determined by the coefficient $\left(g_2/G_2\right)^2$ at the first term in
the right-hand side of Eq. (\ref{er12}) (the common coefficient $\varepsilon$
in Eqs. (\ref{er1})-(\ref{er2}) determines the time scale $\varepsilon t$ of
the evolution of the photon subsystem, as well as the atomic 1st and 3rd levels).
If $|g_2|\ll|g|$, then $|p_{\mathbf{1}n}|^2 \sim (g_2/g)^4$ for $n\ge 2$, whereas
$|p_{\mathbf{3}n}|^2 \sim (g_2/g)^2$ for $n\ge 0$, due to Eq. (\ref{vnb}),
so that the process of photon generation is correlated with the population
of the third level.
In particular, for $g_{2}=0$ (the two-level system)
the coefficient $p_{\mathbf{1}2}$ is not coupled to $p_{\mathbf{1}0}$ in Eq. (\ref{er12})
at the initial moment, so that $p_{\mathbf{1}n}(t) \equiv 0$ for $n\ge 2$,
meaning that  photons
cannot be generated in accordance with \cite{PLA,Roberto}.

In the opposite limit $|g_2|\gg|g|$ we have $\tilde{G}_{n-2}^{2}/G_{n}^2 \approx 1$
and $\tilde{G}_{n}^{2}/G_{n}^2 \approx (n+1)/(n-1)$. Therefore Eqs. (\ref{er1})-(\ref{er2})
become identical with the equations for the photon generation in the empty cavity
(without atoms) given, e.g, by  Eq. (\ref{line1}) with $g= x= 0$.
In this case $|p_{\mathbf{3}n}|^2 \sim \sigma_{33} \sim (g/g_2)^2 \ll 1$.
 It looks like the three-level atom becomes ``invisible'' for the field if $|g_2|\gg |g|$.

Note that the Mandel $Q$-factor is always positive in Fig. \ref{fig2new}, moreover,
it increases with the same increment as the mean photon number for $\varepsilon t>1$, being
always much bigger than $\langle n\rangle$. This means that the photon number
fluctuations are rather strong and the photon statistics is ``hyper-Poissonian'', similar
to the two-level case \cite{Resdyn}.

\subsection{Resonances with creation of two photons}

Looking at equations (\ref{q1})-(\ref{q3}) one can see that choosing certain
nonzero values of the resonance shift $x$
[contained in the functions $\mathcal{W}_{\mathbf{i}}(n)$] one can reduce the
arguments of some exponentials in these equations to zero values, whilst other exponentials
will oscillate with large arguments. In such cases we can perform the RWA and obtain
a smaller set of essential resonantly coupled differential equations.
In particular, simple solutions for $\Delta_2=0$ arise if $2x=\pm G_{2}$.
Then only four probabilities can
be significantly different from zero:
\begin{eqnarray}
|p_{10}|^{2} &\simeq &\cos ^{2}(\nu t),~~|p_{\mathbf{1}2}|^{2}\simeq
\frac{g^2}{G_{2}^2}\sin ^{2}(\nu t),  \label{es1} \\
|p_{\mathbf{2}1}|^{2} &\simeq &\frac{1}{2}{\sin ^{2}(\nu t)},~~|p_{\mathbf{3}%
0}|^{2}\simeq  \frac{g_{2}^2}{2G_{2}^2}\sin ^{2}(\nu
t),  \label{es2}
\end{eqnarray}%
where $\nu \equiv \sqrt{2}qg/G_{2}$ and $G_{2}=\sqrt{2g^{2}+g_{2}^{2}}$. All
other probabilities contain extra factors of the order of
$(\varepsilon/g)^{2}$, so they can be neglected in this approximation. We see that at
most two photons can be generated with a significant probability and the
third level becomes partially populated (if $g_{2}=0$, then the results
coincide with that obtained in \cite{PLA}). A similar effect of an indirect
interaction between different energy levels was discovered in \cite{Tan98},
where the \emph{coupling constants\/} depended on the time-dependent cavity
length $L(t)$ as $g\sim \lbrack L(t)]^{-1/2}$, while the cavity frequency
was supposed to be constant. One more analog is the effect of
\textquotedblleft atomic shaking\textquotedblright\ in cavities with moving
boundaries, studied in \cite{shake}. Recently, an analog of DCE in
three-level systems with time-dependent Rabi frequencies was considered in
\cite{Carus08}. We have verified that simple analytical formulas (\ref{es1}%
)--(\ref{es2}) are in full agreement with exact results obtained by solving
numerically the Schr\"{o}dinger equation (\ref{SEtot})
(the difference turns out to be less than the thickness of lines used in
the plots). 

\begin{figure}[tbh]
\begin{center}
\includegraphics[width=0.49\textwidth]{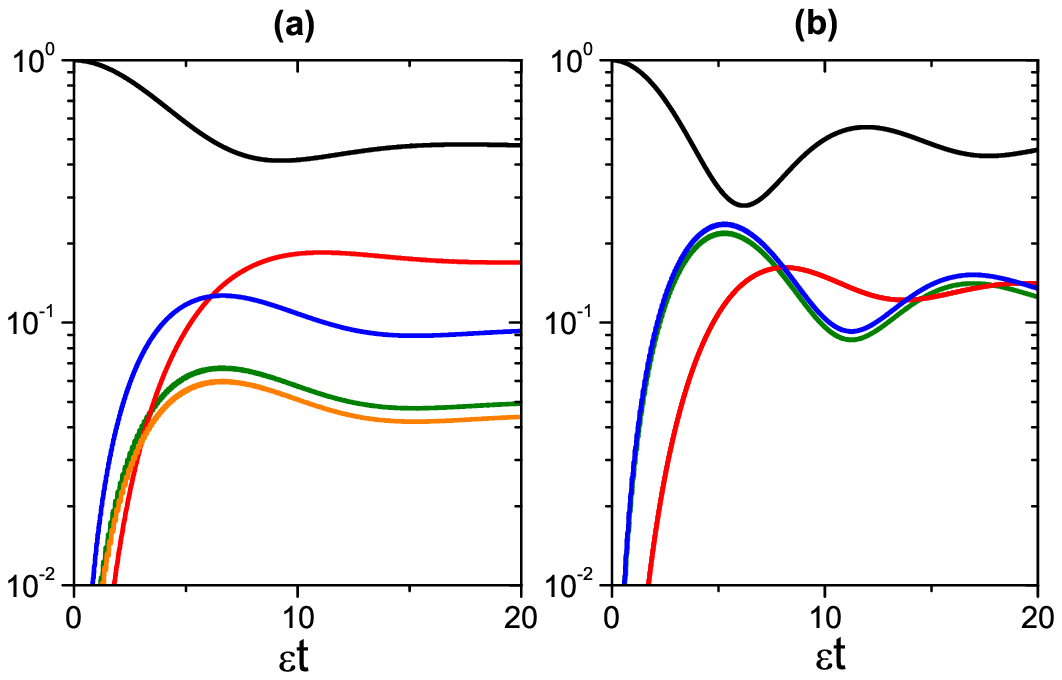} {}
\end{center}
\par
\vspace{-1cm}
\caption{(Color online) Dynamics of nonzero probabilities for the ladder configuration and
nonzero resonance shifts.
The order of curves at $\protect\varepsilon %
t=10$ is as follows (from above). For $2x=G_{2}$ (part a):
$\left\vert p_{\mathbf{1}0}\right\vert ^{2}
$, $\left\vert p_{\mathbf{2}0}\right\vert ^{2}$ (coincident with $\left\vert
p_{\mathbf{1}1}\right\vert ^{2}$), $\left\vert p_{\mathbf{2}1}\right\vert
^{2}$, $\left\vert p_{\mathbf{1}2}\right\vert ^{2}$ and $\left\vert p_{%
\mathbf{3}0}\right\vert ^{3}$.  For $2x=\protect\delta _{2}/2+J$ (part b):
 $\left\vert p_{\mathbf{1}0}\right\vert ^{2}$, $\left\vert p_{\mathbf{%
2}0}\right\vert ^{2}$ (coincident with $\left\vert p_{\mathbf{1}%
1}\right\vert ^{2}$), $\left\vert p_{\mathbf{2}1}\right\vert ^{2}$, $%
\left\vert p_{\mathbf{1}2}\right\vert ^{2}$.
The values of parameters are given in the text.}
\label{fig3new}
\end{figure}

However, simple formulas (\ref{es2}) hold only in the absence of dissipation.
In Fig. \ref{fig3new}a we show numerical results for nonvanishing
probabilities  for parameters: $\Delta _{2}=0$, $g=3\times 10^{-2}$, $%
g_{2}=4\times 10^{-2}$, $\varepsilon =10^{-3}$ and the resonant shift $%
2x=G_{2}$, setting $\lambda =5\times 10^{-4}$ and $\lambda
_{2}=\lambda (g_{2}/g)^{2}$; such a choice agrees with a concrete example of
the transmon multi-level qubit considered in \cite{Koch} (assuming that the
noise couples to the atom via the dipolar interaction \cite{Boise}). One can
see that although no more than two photons can be created from vacuum, there
are no oscillations predicted by  Eq. (\ref{es2}). Moreover, the probabilities
$\left\vert p_{\mathbf{2}0}\right\vert ^{2}$ and $\left\vert
p_{\mathbf{1}1}\right\vert ^{2}$ become different from zero due to the
influence of dissipation.

For the stationary cavity ($q=0$) in the \emph{dispersive case},
$\left\vert \Delta _{2}\right\vert \gg |g_{2}| $,
one can write $p_{\mathbf{3}0}\simeq \left( g_{2}/\Delta
_{2}\right) e^{-i\Delta _{2}t}p_{\mathbf{2}1}$.
 After repeating the same steps as above we find that for $2x=\delta _{2}/2\pm J$
(where $\delta _{2}\equiv g_{2}^{2}/\Delta _{2}$ is the dispersive shift due
to the third level and $J\equiv \sqrt{\delta _{2}^{2}/4+2g^{2}}$) the only
nonzero probabilities are%
\begin{eqnarray}
|p_{\mathbf{1}0}|^{2} &\simeq &\cos ^{2}(q\nu _{\mp }t),\quad |p_{\mathbf{1}%
2}|^{2}\simeq \frac{1}{2}\nu _{\mp }^{2}\sin ^{2}(q\nu _{\mp }t),  \notag \\
|p_{\mathbf{2}1}|^{2} &\simeq &\frac{1}{2}\nu _{\pm }^{2}\sin ^{2}(q\nu
_{\mp }t),\quad \left\vert p_{\mathbf{3}0}\right\vert ^{2}\simeq \frac{
g_{2}^2}{\Delta _{2}^{2}}\,\left\vert p_{\mathbf{2}1}\right\vert ^{2},
\label{pert}
\end{eqnarray}%
where $\nu _{\pm }=\sqrt{1\pm \delta _{2}/(2J)}$. Thus, the rate of photon
generation and the occupation probabilities are influenced by the level $|%
\mathbf{3}\rangle $, although it remains effectively unpopulated for all
times, as was confirmed by numerical simulations. 
In Fig. \ref{fig3new}b we show the probabilities obtained
from the numerical solution of the master equations for $\Delta _{2}=10g_{2}$,
resonance shift $2x=\delta _{2}/2+J$ and other parameters as in Fig. \ref%
{fig3new}a. As expected, the third level is effectively unpopulated and $%
\left\vert p_{21}\right\vert ^{2}$ is slightly bigger than $\left\vert
p_{12}\right\vert ^{2}$, in accordance with the predictions given in Eq. (%
\ref{pert}) for the dissipationless case. Some traces of oscillations are also visible here.
They are more pronounced for smaller values of parameter $\lambda$.

\subsection{Weak field-atom coupling}

For a weak field-atom coupling, $|\varepsilon |\gg |g|$, many photons can be
generated under the resonance condition $x=0$. To calculate the accompanying
atomic dynamics in this case, it is convenient to use the effective
Hamiltonian approach instead of solving the differential equations for the
probability amplitudes \cite{Resdyn}. For this purpose we write the time-dependent
state $|\psi (t)\rangle $ governed by the Hamiltonian $H_{1}$ as%
\begin{equation}
|\psi (t)\rangle =e^{-iH_{1}t}|\psi (0)\rangle =U^{\dagger }\exp \left(
-iH_{ef}t\right) U|\psi (0)\rangle ,
\end{equation}%
where $|\psi (0)\rangle $ is the initial state and we introduced a unitary
operator $U$ to define the effective Hamiltonian $H_{ef}\equiv
UH_{1}U^{\dagger }$. In the resonant regime, $\Delta _{2}=0$, we choose the
transformation \cite{Resact}
\begin{equation}
U=e^{iY},~Y=a^{\dagger }\left( \xi \sigma _{21}+\xi _{2}\sigma _{32}\right)
+a\left( \xi \sigma _{12}+\xi _{2}\sigma _{23}\right) ,
\end{equation}%
where $\xi =2g/\varepsilon \ll 1$ and $\xi _{2}=2g_{2}/\varepsilon \ll 1$.
Then to the second order in $\xi $%
\begin{equation}
H_{ef}=i\theta \left( a^{\dagger 2}-a^{2}\right) +iq\xi \xi _{2}\left(
\sigma _{13}-\sigma _{31}\right) \,,  \label{Hef1}
\end{equation}%
where $\theta \equiv q[1+\xi ^{2}(\sigma _{22}-\sigma _{11})+\xi
_{2}^{2}(\sigma _{33}-\sigma _{22})]$ is an operator with respect to the
atomic basis. This effective Hamiltonian holds approximately for $|g|t\ll 1$%
, so the product $|\varepsilon |t$ can be greater than unity and several
photons can be created from vacuum. We can write%
\begin{eqnarray}
\exp \left( -iH_{ef}t\right) &=&\hat{\Lambda}_{s}\exp \left\{ iqt\left[ \alpha
_{z}\left( \sigma _{33}-\sigma _{11}\right) /2
\right. \right. \nonumber \\ && \left.\left.
+i\xi \xi _{2}(\sigma
_{31}-\sigma _{13})\right] \right\} ,
\label{H-s}
\end{eqnarray}%
where
$\hat{\Lambda}_{v}\equiv \exp [vt(a^{\dagger 2}-a^{2})]$ is the
squeezing operator with nonzero matrix elements in the Fock basis \cite{Puri}
\begin{equation*}
\Lambda _{v}^{(n)}\equiv \langle 2n|\hat{\Lambda}_{v}|0\rangle =\mathcal{C}%
_{v}^{-1/2}(\mathcal{S}_{v}/\mathcal{C}_{v})^{n}\frac{\sqrt{(2n)!}}{2^{n}n!}%
\,.
\end{equation*}%
Here  $\mathcal{C}_{v} \equiv \cosh
\left( 2vt\right) $ and $\mathcal{S}_{v} \equiv \sinh \left( 2vt\right) $.
The operator $\alpha_z$
and index $s$ of the operator $\hat{\Lambda}_{s}$ in (\ref{H-s}) have the form
\begin{eqnarray*}
\alpha _{z} &=&-i\left( \xi ^{2}+\xi _{2}^{2}\right) \left( a^{\dagger
2}-a^{2}\right) , \\
s &=& q\left[ 1+(\xi ^{2}-\xi _{2}^{2})\left( 3\sigma _{22}-1\right) /2\right].
\end{eqnarray*}

After
disentangling the second exponential in Eq. (\ref{H-s}) according to  \cite%
{Puri}, using the property $\hat{\Lambda}_{v}a\hat{\Lambda}_{v}^{\dagger }=%
\mathcal{C}_{v}a-\mathcal{S}_{v}a^{\dagger }$ one can show that for the
initial state $|\psi (0)\rangle =|\mathbf{1},0\rangle $ the occupation
probabilities  to the \emph{second order} in $\xi $ can be written as follows:
\begin{eqnarray*}
\left\vert \langle \mathbf{1},2n|\psi (t)\rangle \right\vert ^{2} &=&\left[
1-2\xi ^{2}\left( n+1\right) \right] \Lambda _{\theta _{1}}^{\left( n\right)
2}
\\ &&
+\Lambda _{\theta _{1}}^{\left( n\right) }\left[ 2\xi ^{2}\left(
2n+1\right) \mathcal{C}_{\theta _{2}}^{-1}\Lambda _{\theta
_{2}}^{(n)}\right.  \\
&&\left. -\left( \xi ^{2}+\xi _{2}^{2}\right) qt\left( 4n\mathcal{S}%
_{2\theta _{1}}^{-1}-\mathcal{S}_{\theta _{1}}\mathcal{C}_{\theta
_{1}}^{-1}\right) \Lambda _{\theta _{1}}^{(n)}\right]
\end{eqnarray*}%
\begin{equation*}
\left\vert \langle \mathbf{2},2n+1|\psi (t)\rangle \right\vert ^{2}=\xi
^{2}\left( 2n+1\right) \left( \mathcal{C}_{\theta _{2}}^{-1}\Lambda _{\theta
_{2}}^{(n)}-\Lambda _{\theta _{1}}^{\left( n\right) }\right) ^{2},
\end{equation*}%
\[
\theta _{1}=q\left[ 1-\left( \xi ^{2}-\xi _{2}^{2}\right) /2\right], ~~
\theta _{2}=q\left[ 1+\left( \xi ^{2}-\xi _{2}^{2}\right) \right] .
\]
The amplitudes related to the third level are very small:
$\left\vert \langle \mathbf{3},2n|\psi (t)\rangle \right\vert
^{2}\propto (\xi \xi _{2})^{2}$
(and other probability amplitudes are exactly zero due to the assumed initial
state $|\mathbf{1},0\rangle $). Therefore, in the resonant regime the third
level is not populated within the time scale $gt\lesssim 1$, but the photon
statistics is nevertheless slightly modified due to the presence of $\xi _{2}
$ in the formulas.We checked these expressions by solving numerically the
Schr\"{o}dinger equation (\ref{SEtot}) and found an excellent agreement. 

\section{V configuration}

Now we repeat the steps of the previous section for the atom-field
interaction Hamiltonian (\ref{HIV})
that describes the atomic configuration depicted in Fig. \ref{f1}b. Using
the transformation $|\Psi (t)\rangle =V(t)|\psi (t)\rangle $ with $V\left(
t\right) =\exp \left[ -it\left( \eta /2\right) \left( n+\sigma _{33}+\sigma
_{22}\right) \right] $ we obtain%
\begin{eqnarray*}
H_{2} &\simeq &(ga\sigma _{21}+g_{3}a\sigma _{31}-iqa^{2}+H.c.) \\
&&-x(n+\sigma _{22}+\sigma _{33})-\Delta _{3}\sigma _{33}.
\end{eqnarray*}%
Only the states $\{|\mathbf{1}%
,2n\rangle$, $|\mathbf{2},2n-1\rangle $, and $|\mathbf{3},2n-1\rangle \}$ are
populated ($n\geq 1$) during the unitary evolution  for the initial state
$|\mathbf{1},0\rangle $. Writing the wavefunction as
\begin{eqnarray*}
|\psi \rangle &=&\sum_{n=0}e^{ixnt}\left( p_{\mathbf{1}n}|\mathbf{1}%
,n\rangle +e^{itx}p_{\mathbf{2}n}|\mathbf{2},n\rangle \right. \\
&&\left. +e^{it(\Delta _{3}+x)}p_{\mathbf{3}n}|\mathbf{3},n\rangle \right) ,
\end{eqnarray*}%
we obtain the equations
\begin{eqnarray*}
\dot{p}_{\mathbf{1}n} &=&-ig\sqrt{n}p_{\mathbf{2}\left( n-1\right) }-ig_{3}%
\sqrt{n}e^{it\Delta _{3}}p_{\mathbf{3}\left( n-1\right) }+\mathcal{W}_{%
\mathbf{1}}(n), \\
\dot{p}_{\mathbf{2}n} &=&-ig\sqrt{n+1}p_{\mathbf{1}\left( n+1\right) }+%
\mathcal{W}_{\mathbf{2}}(n), \\
\dot{p}_{\mathbf{3}n} &=&-ig_{3}\sqrt{n+1}e^{-it\Delta _{3}}p_{\mathbf{1}%
\left( n+1\right) }+\mathcal{W}_{\mathbf{3}}(n).
\end{eqnarray*}%
Their solutions for $\Delta _{3}=q=0$
are
\begin{equation}
p_{\mathbf{1}n}=A_{n}e^{-iG_{n}t}+B_{n}e^{iG_{n}t},  \label{l1}
\end{equation}%
\begin{equation*}
p_{\mathbf{2}\left( n-1\right) }=\frac{g\sqrt{n}}{G_{n}}\left(
A_{n}e^{-iG_{n}t}-B_{n}e^{iG_{n}t}+C_{n}\right) ,
\end{equation*}%
\begin{equation}
p_{\mathbf{3}\left( n-1\right) }=\frac{g_{3}\sqrt{n}}{G_{n}}\left(
A_{n}e^{-iG_{n}t}-B_{n}e^{iG_{n}t}-\frac{g^{2}}{g_{3}^{2}}C_{n}\right) ,
\end{equation}%
where $G_{n}\equiv \sqrt{n(g^{2}+g_{3}^{2})}$. Therefore for the \emph{weak
modulation} ($|\varepsilon |\ll |g|$) the resonances occur for $2x=\pm
G_{n}$ only, resulting in the probabilities%
\begin{equation}
|p_{\mathbf{1}0}|^{2}\simeq \cos ^{2}(qt),\quad |p_{\mathbf{1}2}|^{2}\simeq
\frac{1}{2}\sin ^{2}(qt),
\end{equation}%
\begin{equation}
|p_{\mathbf{2}1}|^{2}\simeq  \frac{g^2}{G_{2}^2}\sin
^{2}(qt),\quad |p_{\mathbf{3}1}|^{2}\simeq \frac{g_{3}^2}{G_{2}^2}
\sin ^{2}(qt).
\end{equation}%
Moreover, since $p_{\mathbf{1}2}$ does not contain the coefficient $C_{2}$
in Eq. (\ref{l1}), the $x=0$ resonance does not appear for the
V configuration. In the dispersive regime, $\left\vert \Delta
_{3}\right\vert \gg |g_{3}|$, one can write $p_{\mathbf{3}1}\simeq (\sqrt{2}%
g_{3}/\Delta _{3})e^{-it\Delta _{3}}p_{\mathbf{1}2}$, where $\delta
_{3}\equiv g_{3}^{2}/\Delta _{3}$, and we find that the resonances occur for
$2x=\delta _{3}\pm J$ with resulting probabilities%
\begin{eqnarray}
|p_{\mathbf{1}0}|^{2} &\simeq &\cos ^{2}(q\nu _{\pm }t),\quad |p_{\mathbf{1}%
2}|^{2}\simeq \frac{1}{2}{\nu _{\pm }^{2}}\sin ^{2}(q\nu _{\pm }t)  \notag \\
|p_{\mathbf{2}1}|^{2} &\simeq &\frac{1}{2}{\nu _{\mp }^{2}}\sin ^{2}(q\nu
_{\pm }t),\quad \left\vert p_{\mathbf{3}1}\right\vert ^{2}\simeq
\frac{2g_{3}^2}{\Delta _{3}^{2}}\,\left\vert p_{\mathbf{1}2}\right\vert ^{2}
\label{pert1}
\end{eqnarray}%
where $J\equiv \sqrt{\delta _{3}^{2}+2g^{2}}$ and $\nu _{\pm }=\sqrt{1\pm
\delta _{3}/J}$. Notice that the expressions (\ref{pert1}) are slightly
different from the corresponding expressions (\ref{pert}) for the
ladder configuration.

For the \emph{strong modulation}, $|\varepsilon |\gg |g|$, we perform the
transformation
\begin{equation}
U=e^{iY},~Y=a^{\dagger }\left( \xi \sigma _{21}+\xi _{3}\sigma _{31}\right)
+a\left( \xi \sigma _{12}+\xi _{3}\sigma _{13}\right) ,
\end{equation}%
where $\xi =2g/\varepsilon \ll 1$ and $\xi _{3}=2g_{3}/\varepsilon \ll 1,$
to obtain the effective Hamiltonian in the resonant regime
(valid for $t\ll |g|^{-1}$ to the second order in $\xi$)%
\begin{equation}
H_{ef}=i[\theta +q\xi \xi _{3}\left( \sigma _{23}+\sigma _{32}\right)
]\left( a^{\dagger 2}-a^{2}\right) ,  \label{lars}
\end{equation}%
where $\theta \equiv q[1+\xi ^{2}(\sigma _{22}-\sigma _{11})+\xi
_{3}^{2}(\sigma _{33}-\sigma _{11})]$. After disentangling $\exp (-iH_{ef}t)$
we get the following nonvanishing probabilities for the initial state $|%
\mathbf{1},0\rangle $:
\begin{eqnarray*}
\left\vert \langle \mathbf{1},2n|\psi (t)\rangle \right\vert ^{2} &=&\left[
1-2\left( \xi ^{2}+\xi _{3}^{2}\right) \left( n+1\right) \right] \Lambda
_{\theta _{1}}^{(n)2} \\
&&+2\left( 2n+1\right) \left( \xi ^{2}+\xi _{3}^{2}\right) \mathcal{C}%
_{\theta _{2}}^{-1}\Lambda _{\theta _{2}}^{(n)}\Lambda _{\theta _{1}}^{(n)},
\end{eqnarray*}%
\begin{equation*}
\left\vert \langle \mathbf{2},2n+1|\psi (t)\rangle \right\vert ^{2}=\xi
^{2}\left( 2n+1\right) \left( \mathcal{C}_{\theta _{2}}^{-1}\Lambda _{\theta
_{2}}^{(n)}-\Lambda _{\theta _{1}}^{(n)}\right) ^{2},
\end{equation*}%
\begin{equation*}
\left\vert \langle \mathbf{3},2n+1|\psi (t)\rangle \right\vert ^{2}=\xi
_{3}^{2}\left( 2n+1\right) \left( \mathcal{C}_{\theta _{2}}^{-1}\Lambda
_{\theta _{2}}^{(n)}-\Lambda _{\theta _{1}}^{(n)}\right) ^{2},
\end{equation*}%
where $\theta _{1}=q(1-\xi ^{2}-\xi _{3}^{2})$ and $\theta _{2}=q[1+(\xi
^{2}+\xi _{3}^{2})/2]$. As expected for the resonant regime, the third level
can be substantially populated in this case, and the photon statistics is
severely modified as compared to the scenario of resonant two-level atom.

\begin{figure}[tbh]
\begin{center}
\includegraphics[width=0.49\textwidth]{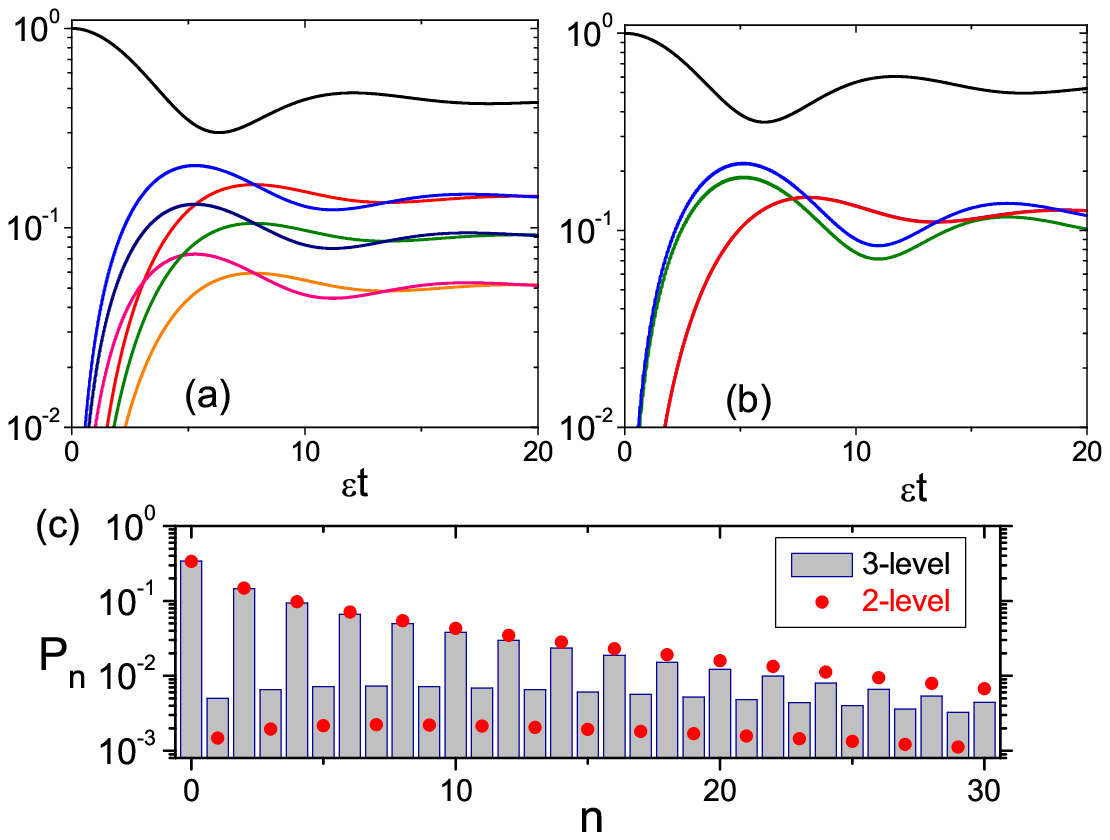} {}
\end{center}
\par
\vspace{-1cm}
\caption{(Color online) Atom-field dynamics for the V configuration and
different resonance shifts. The order of curves at $\protect\varepsilon %
t=10$ is as follows (from above).
\textbf{a}) For $2x=G_{2}$:  $|p_{\mathbf{1}0}|^2$, $|p_{\mathbf{1}1}|^2$,
$|p_{\mathbf{1}2}|^2$, $|p_{\mathbf{3}0}|^2$, $|p_{\mathbf{3}1}|^2$,
$|p_{\mathbf{2}0}|^2$, $|p_{\mathbf{2}1}|^2$.
\textbf{b}) For $2x=\protect\delta _{3}+J$:  $|p_{\mathbf{1}0}|^2$,
$|p_{\mathbf{1}1}|^2$ (coincident with $|p_{\mathbf{2}0}|^2$), $|p_{\mathbf{2}1}|^2$,
$|p_{\mathbf{1}2}|^2$. \textbf{c}) Photon number distribution for $x=0$ and $\protect%
\varepsilon t=3.5$ in the strong modulation regime and $\Delta _{3}=-4g_{3}$
(without damping, $\protect\lambda =\protect\lambda _{3}=0$), compared to
the two-level atom case ($g_{3}=0$). The values of other parameters are given in the text.}
\label{f3}
\end{figure}

In Fig. \ref{f3} we illustrate the exact dynamics for the V configuration.
Fig. \ref{f3}a shows the behavior of probabilities in the resonant regime ($%
\Delta _{3}=0$) and weak modulation ($\varepsilon =10^{-3}$) for $2x=G_{2}$,
while in Fig. \ref{f3}b we consider the dispersive regime ($\Delta
_{3}=-12g_{3}$) and the resonance shift $2x=\delta _{3}+J$. We included
atomic damping by means of the master equation (\ref{meq}),
replacing the term $\lambda _{2}\mathfrak{D}[\sigma _{23}]\rho $
by $\lambda _{3}\mathfrak{D}[\sigma _{13}]\rho $, where $\lambda _{3}$ is
that for the transition $|\mathbf{3}\rangle \rightarrow |\mathbf{1}\rangle $
[other parameters are: $g=3\times 10^{-2}$, $g_{3}=4\times 10^{-2}$, $%
\lambda =5\times 10^{-4}$, $\lambda _{3}=\lambda (g_{3}/g)^{2}$]. In both
cases at most two photons are created as predicted analytically in the
absence of damping, and in the dispersive case $\left\vert p_{\mathbf{2}%
1}\right\vert ^{2}$ lies slightly above $\left\vert p_{\mathbf{1}%
2}\right\vert ^{2}$ in accordance with Eq. (\ref{pert1}). In Fig. \ref{f3}c
we show the photon distribution (obtained by
tracing out the atomic degrees of freedom) in the absence of damping
 for $x=0$ in the strong
modulation regime for parameters $\varepsilon =10^{-2}$, $g=5\times 10^{-4}$%
, $g_{3}=8\times 10^{-4}$ and $\Delta _{3}=-4g_{3}$ (so the third level is
neither in resonant nor in dispersive regime). For comparison we show the
photon number distribution in the absence of the third level ($g_{3}=0$) to
emphasize that the photon statistics is substantially modified due to the
interaction with the third level.

\section{Conclusions}

We presented results of exact numerical calculations
 for the atom-field dynamics
when a three-level atom (see Fig. \ref{f1}) interacts with a single cavity
field mode whose vacuum state is being amplified via the dynamical Casimir
effect. In some cases we succeeded to find  simple analytical expressions
explaining these results.
This study is relevant since the actual atoms in cavity QED and artificial atoms
in circuit QED are indeed multi-level systems. We found that the third level
modifies the resonance frequencies as compared to the two-level case, and
the dynamical behavior may be drastically different from the cases of
atomless cavity or a two-level atom.
The results obtained
might be useful for the design of schemes aimed at the detection of the Casimir
photons by measuring the occupancies of different atomic levels.
For instance, the modulation frequency
equal to twice the unperturbed cavity frequency does lead to photon
generation from vacuum and occupation of the third level in the
ladder configuration, whereas this modulation frequency is forbidden in the
case of two-level resonant atom or V configuration. This could facilitate
the experiment, because there is no need in such a case to adjust the resonance
frequency shift, knowing that the main resonance must happen exactly at twice
the frequency of the unperturbed cavity mode, while
the occupancy of the third level in the ladder configuration
can serve as a witness of the photon
generation, because whenever the photons are generated the third level becomes populated.
In any case, the inclusion of the third level provides an opportunity
for observing a rich dynamical behavior.
Also, the three-level schemes can be useful for the creation
of different entangled states between the field and atoms
(whereas by using post-selection methods based on detecting the
atomic state, novel cavity field states could be engineered \cite{Resdyn}).

\begin{acknowledgments}
A.V.D. acknowledges the partial support of DPP/UnB. V.V.D. acknowledges the partial support of CNPq (Brazilian agency).
\end{acknowledgments}

\end{document}